\documentclass[a4paper,12pt]{article}
\pdfoutput = 1

\usepackage{ucs}

\usepackage{amssymb}
\usepackage{mathrsfs}
\usepackage{amsmath,amsfonts,amssymb,amsthm,mathtools}
\usepackage{multirow}
\usepackage{tikz}

\usepackage{graphicx}
\usepackage{bm}
\usepackage[left=12mm,right=15mm,top=20mm,bottom=30mm,bindingoffset=1.5cm]
           {geometry}

\usepackage{indentfirst}           
\usepackage{epstopdf}
\makeatletter
\def\@xfootnote[#1]{%
  \protected@xdef\@thefnmark{#1}%
  \@footnotemark\@footnotetext}
\makeatother

\newcommand \mm{{\mu^6_{\Lambda}} }

\title{\bf Dimension-seven operator contribution to the
    top quark  anomalous interactions }
\author{V.V.~Denisov$^{1}$,
  S.R.~Slabospitskii${}^{2}$  }
\date{}
\begin{document}

\maketitle

 {\small
   {\em
     \begin{center}
       ${}^1$ Moscow Institute of Physics and Technology, \\
     Dolgoprudny, Moscow Region, Russia\\ 
    ${}^2$ NRC ``Kurchatov Institute'' - IHEP, \\
Protvino, Moscow Region, Russia 
\\
\end{center}
E--mail:~~Vladislav.Denisov@ihep.ru, ~~Sergei.Slabospitskii@ihep.ru \\ \\
}
}

\vspace{5mm}

\vspace{50 pt}

\begin{abstract}
  The contribution of dimension-seven operators to anomalous
  FCNC-interactions of the $t$-quarks with a photon and a gluon
  is considered.
 The phenomenological
 Lagrangian and Feynman rules are derived.
 There are evaluated the expressions for
  the widths FCNC-decays of the $t$-quark into light quark and two photons,
  two gluons, a photon
and a gluon, and three quarks.
\end{abstract}

\normalsize
\thispagestyle{empty} 

\newpage
\pagenumbering{arabic}

\section{Introduction}

At present, it is not known what type of New Physics (NP)
beyond  the Standard Model framework 
will be responsible for possible deviations from
predictions of the Standard Model (SM).
 Multiple scenarios of SM extensions 
  lead to different predictions in the  $t$-quark sector with their own
  a specific set of types of interactions and parameters (coupling 
  constants, the masses of new objects). At the same time,
  different scenarios often
  predict the same or very similar effects, leading 
  to processes with identical final states.

  To describe the various anomalous interactions of $t$-quarks
  it is widely used a
  universal approach based on the formalism of an effective field
theory~\cite{Buchmuller:1985jz}.
In this approach, the anomalous interactions of $t$-quarks are described
by a model-independent manner through the use of
an effective (phenomenological)
Lagrangian~\cite{Beneke, AguilarSaavedra:2008zc, Grzadkowski:2010es, eff}.
This Lagrangian must be gauge-invariant with respect 
to the SM gauge group
(otherwise, the input anomalous interactions immediately
would lead to contradictions
with modern precision measurements) and consists of the terms 
 with an increasing dimension,
suppressed by increasingly higher degrees of the NP-scale.
Such a Lagrangian of the anomalous $t$-quark interactions 
can be presented in the following
form~\cite{Buchmuller:1985jz, AguilarSaavedra:2008zc,
  Grzadkowski:2010es}:
\begin{eqnarray}
  {\mathcal L}_{QFT} =  {\mathcal L}_{SM} +
  \kappa_4 \bar{\psi}_q \hat{O}^{(4)} \psi_t +
  \frac{\kappa_5}{\Lambda}\bar{\psi}_q \hat{O}^{(5)} \psi_t +
  \frac{\kappa_6}{\Lambda^2}\bar{\psi}_q \hat{O}^{(6)} \psi_t+ \cdots
\label{lagrangian-1}
\end{eqnarray}
where ${\mathcal L}_{SM}$ is the SM Lagrangian
(see, e.g.~\cite{Borodulin:2017pwh}),
$\Lambda$ is new physics mass scale,
$\kappa_i$ are the anomalous couplings.  

The promising directions of searching for NP in the $t$-quark sector
are processes with
the Flavour changing neutral currents (FCNC) 
\begin{equation*}
t \rightarrow \gamma(g, \, Z) + c(u)
\end{equation*}
Within the SM framework such processes are highly
suppressed (due to loop diagrams)~\cite{Beneke}:
$
B (t \rightarrow q\;\gamma/g/Z) < \mathcal{O}(10^{-11}\div10^{-13})$,
which makes them practically impossible to observe in the experiment.
Thus, the experimental observation 
 of the $t$-quark FCNC interactions
$t$-quarks will unambiguously indicate
 the existence of NP beyond the SM.

The FCNC processes with the dimension-5 and 6 operators
were analyzed in detail earlier
and there were built
expressions for all
possible FCNC interactions of $t$-quarks
(see, for example,~\cite{AguilarSaavedra:2008zc, Grzadkowski:2010es}).

In this article we consider the role of the dimension-seven operator
contributions to anomalous 
FCNC-interaction of $t$-quarks with a photon
and gluon ($ t \, \gamma \, q$ and $t \, g \, q)$:
\begin{eqnarray*}
  \mathscr{L}^{(7)}_{FCNC} = \cfrac{\kappa}{\Lambda^3}\bar{\psi}_{q}
  \hat{O}^{(7)}\psi_{t}, \;
  q = u,\;c 
\end{eqnarray*}
In this article, the
upper $q$-quark in the $(tVq)$ interaction will be denoted by the
symbol $u$.
Therefore,   all the results will remain the same for $u$- or $c$-quarks. 

\section{Anomalous FCNC interaction Lagrangian }
In this article, when constructing the operators $\hat{O}^{(n)} $
there are considered the interactions with only
  {\bf one} external massless
   gauge boson $V$ ($V = \gamma, \, g $).
Due to a gauge invariance, the interacting boson enters into
  operators
 in the form of the field strength tensor
  $\mathcal {F}^{\mu \nu} $
  (and the dual tensor $\widetilde {\mathcal {F}}_{\mu \nu} $):
  \begin{eqnarray}
     \mathscr{L}_{FCNC}
  &=& \frac{1}{\Lambda^3}\bar{\psi}_{u} \kappa_1 \, 
  \hat{W}^{(7)\; \mu\nu}\psi_{t} \mathcal{F}_{\mu\nu}
  + \frac{1}{\Lambda^3}\bar{\psi}_{u} \, \kappa_2
  \hat{W}^{(7)\; \mu\nu}\psi_{t} \widetilde{\mathcal{F}}_{\mu\nu}
  \label{lagrangian-2} \\
 \widetilde{\mathcal{F}}_{\mu\nu} &\equiv &
 \varepsilon_{\mu \nu \alpha \beta} \mathcal{F}^{\alpha \beta };
 \quad \kappa_i = \xi_i \, + \, \zeta_i \, \gamma^5 
   \nonumber
   \\
 \mathcal{F}_{\mu \nu}^{(qed)} &=&
     F_{\mu\nu} = \partial_{\mu}A_{\nu}
  - \partial_{\nu}A_{\mu}  \nonumber \\
    \mathcal{F}_{\mu \nu}^{(qcd)} &=& G^a_{\mu\nu}= \partial_{\mu}B^a_{\nu} -
    \partial_{\nu}B^a_{\mu}+if^{abc}t^c B_{\mu}^b B_{\nu}^c \nonumber     
 \nonumber 
 \end{eqnarray}
  where $A_{\mu}, B^a_{\mu}$ are the photon and gluon fields,
  $\xi_i$ и $\zeta_i$ - are the anomalous couplings
  (in the general case, the complex numbers).
  
  In the general case, the operator $\hat{W}^{(n)}$
  is constructed from the Dirac, Gell-Man
  and covariant derivatives
  $D^{\mu}, \; D^{* \, \mu}$~\cite{Borodulin:2017pwh}:
\begin{eqnarray*}
  &&\overrightarrow{D}^{\mu} =
  D^{\mu} = \overrightarrow{\partial^{\mu}} -
  i e_qA^{\mu} - ig_st^a G_{a}^{\mu}\\
  &&\overleftarrow{D}^{\* \mu} =
  D^{*\; \mu} = \overleftarrow{\partial^{\mu}}
  + i e_q A^{\mu} + ig_st^aG_{a}^{\mu}
\end{eqnarray*}
where $ e_q $ is the electric charge of the quark, $ g_s $
is the constant of the strong
interactions, and $t^a $ - Gell-Mann matrices.
The covariant derivative $ D^{\mu} $ acts
on the spinor $\psi_t $, and $ D^{* \; \mu} $ - to the antispinor
$\bar{\psi}_u $.
We note that,
because of covariant derivatives, in Lagrangians 
there are terms,
describing the interaction with one, two and three bosons.
In this article we present an explicit form of expressions only
for the interaction
with one and two bosons.

{\bf Second} restriction on the form of the operators $\hat{W}^{(n)}$
consists in the following.
When constructing such operators, it is assumed that
covariant derivatives act {\bf only} on quarks (spinors).
In this case, they must have convolution by indices with a
strength tensor.
This assumption restricts the type of the operators $\hat{W}^{(n)} $.
Indeed, let us consider the contribution of the dimension-seven operator:
\[
\bar{\psi}_u \hat{D^{*}} \gamma^{\mu}  D^{\nu} \psi_t \mathcal{F}_{\mu \nu},
\;\;\; \hat{D^{*}} =  
  \overleftarrow{D}^{\* \alpha} \gamma^{\alpha} 
\]
In this case, the action of the derivative $\hat{D^{*}}$ 
(which is not related to
interacting boson), can be treated, for example, as a form-factor:
\begin{eqnarray*} 
  & &\frac{\kappa^{(7)}}{\Lambda^3} \bar{\psi}_u \hat{D^{*}} \gamma^{\mu}
  D^{\nu} \psi_t
  \mathcal{F}_{\mu \nu} =  \frac{\kappa^{(7)}}{\Lambda^2}
  \bar{\psi}_u \left(
  \frac{\overleftarrow{D}^{\* \alpha} \gamma^{\alpha}}{\Lambda}  \right)
  \gamma^{\mu}  D^{\nu}
  \psi_t \mathcal{F}_{\mu \nu};  \;\;\; 
 \frac{\kappa^{(7)}}{\Lambda^2}
  \bar{\psi}_u  \left( 
  \frac{\overleftarrow{D}^{\* \alpha}\gamma^{\alpha}}{\Lambda} \right) 
  \Longrightarrow  \frac{\tilde{\kappa}(q^2)}{\Lambda^2} \bar{\psi}_u \\
&& \Longrightarrow \, \frac{\kappa^{(7)}}{\Lambda^3} \bar{\psi}_u \hat{D^{*}}
  \gamma^{\mu}  D^{\nu} \psi_t
  \mathcal{F}_{\mu \nu} \simeq 
  \frac{\tilde{\kappa}(q^2)}{\Lambda^2}
  \bar{\psi}_u   \gamma^{\mu}  D^{\nu} \psi_t \mathcal{F}_{\mu \nu}
\end{eqnarray*}
The last expression is, in fact, represented by the contribution of an
operator of dimension-six!
Thus, taking into account these two assumptions
(on the interaction with only one boson and
second, described above) the phenomenological Lagrangian
of FCNC
interactions can be constructed only
from the operators of dimensions-five, six and seven.

\subsection{The anomalous interaction with a photon}
Dimension-seven operator for anomalous $t$-quark FCNC-interaction
with a photon comprises  four gauge-invariant terms:
\begin{eqnarray*}
  &&  \hat{W}^{(7) \; \mu \nu}_{\gamma} \; : \;   D^{\mu} D^{\nu}, \quad
  D^{* \; \mu} D^{* \; \nu}, \quad  D^{* \; \mu}D^{\nu},
  \quad  D^{\nu} D^{* \; \mu}
 \end{eqnarray*}
Taking into account the antisymmetry of the
$F_{\mu \nu}$ the contributions of the first and second terms
are equal to each other, and the last two are related by the relation:
\begin{eqnarray*}
  &&    D^{\mu}D^{\nu} = D^{* \, \mu} D^{*\, \nu} =
  -\frac{1}{2} \Delta^{\mu \nu} \nonumber \\
  &&  D^{\nu} D^{* \, \mu} = D^{* \, \mu} D^{\nu} \,
  \, - \Delta^{\mu \nu}  \nonumber \\
  && \Delta^{\mu \nu} = i e_q \, F^{\mu \nu} + ig_s
  t^a G_a^{\mu \nu} 
\end{eqnarray*}
Therefore, the operator $\hat{W}^{(7) \; \mu \nu}_{\gamma} $
comprises only two independent structures ($D^{* \, \mu} D^{\nu} $
and $\Delta^{\mu \nu} $). The Lagrangian of $t$-quark
FCNC-interaction with a photon has the form:
\begin{eqnarray} 
&&  \mathscr{L}^{(7)}_{FCNC}(t \, \gamma q)
  = \frac{e_q}{\Lambda^3}\bar{\psi}_{u}
  \hat{W}^{\mu\nu}\psi_{t} F_{\mu\nu}
  + \frac{e_q}{\Lambda^3}\bar{\psi}_{u}
  \hat{W}^{\mu\nu}_D\psi_{t} \widetilde{F}_{\mu\nu} \label{qed-3} \\
  && \hat{W}^{\mu\nu} =  \kappa_{1}  D^{*\, \mu}D^{\nu}
  - \kappa_2 \frac{\Delta^{\mu \nu}}{2}; \quad 
  \hat{W}^{\mu\nu}_D =  \kappa_{3}  D^{*\, \mu}D^{\nu}
  - \kappa_4 \frac{\Delta^{\mu \nu}}{2} 
  \nonumber \\
  && \kappa_{i} =  \xi^{\gamma}_i+\zeta^{\gamma}_i\gamma^5 
  \;\;\; \nonumber
\end{eqnarray}
where $\xi^{\gamma}_i,\;\zeta^{\gamma}_i$ are the anomalous
couplings.  

In what follows, the common  numerical factors
(of the type $\pm 1 $, 1/2, $ i, ... $) are included in the anomalous
couplings.
We note that in the Lagrangian the operators
$D^{* \, \mu} D^{\nu} \times
\widetilde{F}_{\mu \nu} $
and $\Delta^{\mu \nu} \times
\widetilde {F}_{\mu \nu} $ result in the same expressions for
terms describing the interaction with one or two
bosons. Therefore, in this article we rely on
\[
\kappa_4 = \kappa_3
\]
Omitting the trivial computations and using the the momentum 
representation,
for each terms from~$\hat{W}^{(7)}$ and $\hat{W}^{(7)}_D$
we have
\begin{eqnarray*}
\begin{array}{lcl}
\bar{\psi_u}D^{* \, \mu}D^{\nu}\psi_t F_{\mu\nu} & \to & 
\bar{u}_u \kappa_1  [\hat{w}_1+\hat{X}_2+\hat{U}_2 +\hat{V}_1(3)]u_t
\\
\bar{\psi_u} \Delta^{\mu \nu} \psi_t F^{\mu\nu} & \to & 
\bar{u}_u \kappa_2  [2X_1 + U_1 + V_1(3)]u_t
\\[2mm]
\bar{\psi_u} D^{* \,\mu} D^{\nu} \psi_t \widetilde{F}_{\mu\nu}  =
\bar{\psi_u} \Delta^{\mu \nu} \psi_t \widetilde{F}_{\mu\nu} 
& \to & 
\bar{u}_u \kappa_3  [2\hat{X}_3 + \hat{U}_3 + \hat{V}_2(3)]u_t
\end{array} 
\end{eqnarray*}
where $\bar{u}_u $ and $u_t$ are the spinors of the light $u$ and
$t$-quarks with momenta
$p_2$ and $p_1$, respectively. The expressions $\hat{V}_i (3), \, i = 1,2 $
describe interactions with three
bosons. The explicit form of which is not given in this article.
For the rest, we have (below, each expression contains
the total coefficient $\Lambda^{-3} $):
\begin{eqnarray} 
  \left.
 \begin{array}{lcl}
w_1 &=&e_q p_1^{\mu}p_2^{\nu}(q^{\mu}g^{\nu\alpha}-q^{\nu}g^{\mu\alpha})A^{\alpha}
\\
X_1 & = & e_q^2\left[(q_1q_2)g^{\alpha\beta}-q_1^{\beta}q_2^{\alpha}\right]
A_1^{\alpha}A_2^{\beta}
\\ 
X_2 & = & e_q^2\left[(q_1+q_2)^2 g^{\alpha\beta} 
  -q_2^{\alpha}(q_1+q_2)^{\beta} - q_1^{\beta}(q_1+q_2)^{\alpha}
  \right]A_1^{\alpha}A_2^{\beta}
\\ 
X_3 &= & e_q^2 \varepsilon^{\mu\nu\alpha\beta} q_1^{\mu}q_2^{\nu} A_1^{\alpha}
A_2^{\beta}
\\
U_1 & = & e_qg_s t^a
\left[(q_1q_2)g^{\alpha\beta}-q_1^{\beta}q_2^{\alpha}\right]
A^{\alpha}(q_1) B_{a}^{\beta}(q_2)
\\
U_2 & = & e_qg_st^a (q_1+q_2)^{\lambda}(q_1^{\lambda} g^{\alpha\beta}
- q_1^{\beta}g^{\lambda\alpha})A^{\alpha}(q_1)B_{a}^{\beta}(q_2)
\\
U_3 & = & e_q g_st^a \varepsilon^{\mu\nu\alpha\beta}
q_1^{\mu}q_2^{\nu}A^{\alpha}(q_1)B_{a}^{\beta}(q_2)
 \end{array}
 \right\} \label{qed-5}
\end{eqnarray}
here $q_1$ и $q_2$ are the momenta of the bosons.
The corresponding Feynman rules are given in the Appendix.

\subsection{The anomalous interaction with a gluon}
Dimension-seven operator for anomalous $t$-quark FCNC-interaction
with a gluon comprises three gauge-invariant terms:
\begin{eqnarray*}
  &&  \hat{W}^{a\;\mu \nu} \; : \;   D^{*\, \mu} t^a D^{\nu}, \quad
  t^a D^{\mu} D^{\nu}, \quad  D^{*\, \mu}D^{*\, \nu} t^a 
\end{eqnarray*}
Also, as in the case of interaction with a photon,
the contributions of operators
$t^a D^{\mu} D^{\nu}$ and $D^{* \, \mu} D^{* \, \nu} t^a $
are equal to each other. Thus, the  Lagrangian of dimension-seven,
describing the interaction with the gluon, has the form:
\begin{eqnarray}
 &&  \mathscr{L}^{(7)}_{FCNC}(t \, g \, q)
  = \frac{g_s}{\Lambda^3}\bar{\psi}_{u}
  \hat{W}^{a \, \mu\nu}\psi_{t} G^a_{\mu\nu}
  + \frac{g_s}{\Lambda^3}\bar{\psi}_{u}
  \hat{W}^{a\, \mu\nu}_D\psi_{t} \widetilde{G}^a_{\mu\nu} \label{qcd-1} \\
  && \hat{W}^{ a\,  \mu\nu} =  \lambda_1  D^{*\, \mu} \, t^a \, D^{\nu}
  - \lambda_2 \, t^a \frac{\Delta^{\mu \nu}}{2};
  \;\;
  \hat{W}^{ a\,  \mu\nu}_D =  \lambda_3  D^{*\, \mu} \, t^a \, D^{\nu}
  - \lambda_4 \, t^a \frac{\Delta^{\mu \nu}}{2}
 \nonumber
  \\
  &&\lambda_{i} =  \xi^g_i+\zeta^g_i\gamma^5
\nonumber
\end{eqnarray}
where $\xi^g_i, \; \zeta^g_i $ are anomalous couplings, 
$\bar{\psi}_u $ and $\psi_t $ are spinors describing the light
$u$ and $t$-quarks, the strength tensor $ G_a^{\mu \nu} $ is defined
above~(\ref{lagrangian-2}).

After transition into momentum
space for each term in the Lagrangian~(\ref{qcd-1}) we get: 
\begin{eqnarray*}
 \begin{array}{lcl}
 \bar{\psi}_u D^{*\, \mu} t^a D^{\nu} \psi_t G^a_{\mu \nu} & \to & 
 \;\; \bar{u}_u  \lambda_1 \left[w^g_1 + U_0  + Y_2  + V_{3,4} \right]u_t
 \\
\bar{\psi}_u t^a D^{\mu} D^{\nu}  \psi_t  G^a_{\mu \nu}  & \to &
\;\; \bar{u}_u  \lambda_2  \left[U_1  + Y_1 +V_{3,4}\right] u_t
\\
\bar{\psi}_u  D^{* \mu} t^a D^{\nu} \psi_t  \widetilde{G}^a{_\mu \nu } & \to & 
\;2\bar{u}_u  \lambda_3
\left[  U_3  + Y_4  + V_{3,4} \right]   u_t
\\
\bar{\psi}_u  t^ a D^{\mu} D^{\nu} \psi_t
\widetilde{G}^a_{\mu \nu} 
& \to &
-2\, \bar{u}_u \lambda_4 \left[ U_3 +   Y_3 + V_{3,4} \right]  u_t
\end{array}
\end{eqnarray*}
where $\bar{u}_u $ and $ u_t $ are spinors,
describing a light $u$ and $t$-quarks
with momenta $ p_2 $ and $ p_1 $, respectively.
Values of $ V_{3,4} $ describe interactions with 3 and 4
bosons. The explicit form of which is not given in this article.
For the rest, we have (below, each expression contains
the total coefficient $\Lambda^{- 3} $):
\begin{eqnarray} 
\left.
\begin{array}{lcl} 
  w^g_1 &= &g_s t^a p_1^{\mu} p_2^{\nu}
  (q^{\mu}g^{\nu\alpha}-q^{\nu}g^{\mu\alpha}) B^{\alpha}_{a}
  \\
  U_0 & = & e_q g_st^a \left[ ((q_1+q_2)q_2) g^{\alpha \beta} -
    q_2^{\alpha} (q_1+q_2)^{\beta} \right] 
  A^{\alpha}_1 B_{a\, 2}^{\beta} 
\\
 U_1 & = & e_q g_s t^a \left[ (q_1q_2) g^{\alpha \beta}
   - q_2^{\alpha} q_1^{\beta} \right]  A^{\alpha}(q_1) B_{a}^{\beta}(q_2)
 \\
 U_3 & = & e_q g_st^a\varepsilon^{\mu\nu\alpha\beta}
 q_1^{\mu}q_2^{\nu} A^{\alpha}_1 B_{a \,2}^{\beta}
 \\
 Y_2 &  = & g_s^2 \left\{
 t^at^b\left[((p_1q_2)-(p_2q_1))g^{\alpha\beta}
   - p_2^{\alpha}p_1^{\beta} + p_1^{\alpha}p_2^{\beta} - q_2^{\alpha}p_1^{\beta}
   + p_2^{\alpha}q_1^{\beta}\right] \right. \\
 &+ & \;\;\;\; \; t^bt^a\left.\left[
   ((p_1q_1)-(p_2q_2))g^{\alpha\beta}+p_2^{\alpha}p_1^{\beta}
   -p_1^{\alpha}p_2^{\beta}+q_2^{\alpha}p_2^{\beta}
   -p_1^{\alpha}q_1^{\beta}\right]\right\} B_{1\;a}^{\alpha} B_{2\;b}^{\beta}
 \\
 Y_1 & = & g_s^2(\frac{1}{3}\delta^{ab}+d^{abc}t^c)
 [(q_1q_2)g^{\alpha\beta}-q_2^{\alpha}q_1^{\beta}]
 B_{1\;a}^{\alpha} B_{2\;b}^{\beta}
 \\ 
 Y_3 & = & g_s^2
 \left[ \left (\frac{1}{3}\delta^{ab}+d^{abc}t^c\right)
    q_1^{\mu} q_2^{\nu} + 2i t^k f^{kab} p_2^{\mu} p_1^{\nu} \right]
 \varepsilon^{\mu\nu\alpha\beta} B_{1\;a}^{\alpha} B_{2\;b}^{\beta}
 \\ 
Y_4 & = & g_s^2
  \left (\frac{1}{3}\delta^{ab}+d^{abc}t^c\right)
  \varepsilon^{\mu\nu\alpha\beta}   q_1^{\mu} q_2^{\nu} 
 B_{1\;a}^{\alpha} B_{2\;b}^{\beta}
\end{array}
\right\}
\label{qcd-4}
\end{eqnarray}
here $q_1$ и $q_2$ are the momenta of the bosons.
The corresponding Feynman rules are given in the Appendix.


\section{$t$-quark decay widths}
We note that the amplitudes ($T$) containing
the anomalous interaction vertex
with one real boson (photon or gluon), are always equal to zero.
Indeed, taking into account the law of conservation of momentum and
choosing a calibration,
  which ensures the Lorentz condition ($ (qV) = 0 $), for such vertices
  we obtain:
\begin{eqnarray*} 
&& T \propto w_1 =  p_1^{\mu} p_2^{\nu}
 (q^{\mu}g^{\nu\alpha}-q^{\nu}g^{\mu\alpha}) V^{\alpha}; \;\; 
 V^{\alpha} = A^{\alpha}, \; B_a^{\alpha}, \;\;
 p_1 = p_2 + q, \;\; (qV) = 0, \;\;q^2 = 0 \\
  &\to&   
 w_1 = \left[(p_1 q) p_2^{\alpha} - (p_2q) p_1^{\alpha} \right] V^{\alpha}
 = \left[(p_2 q) p_2^{\alpha} - q^2 p_2^{\alpha} - (p_2q)q^{\alpha} -
   (p_2q)p_2^{\alpha} \right] V^{\alpha}=0
\end{eqnarray*}
Therefore, in contrast to operators of dimension-5 and 6,
for the considered interaction due to dimension-seven operators
in the lowest order
perturbation theory the $t$-quark can decay into the following
three-body channels:
\begin{eqnarray}
  \left. 
  \begin{array}{lll}
   t \rightarrow u \, \gamma \, \gamma,  
  & t \rightarrow u \, \gamma \, g,  
  & t \rightarrow u \, g \, g   \\
  t \rightarrow u \, q \, \bar{q}; \quad q \ne u, 
  & t \rightarrow u \, \bar{u} \, u &
  \end{array} \right. \label{channels}
\end{eqnarray}
The diagrams describing these processes are presented below.
\begin{figure}[h]
  \begin{center}
 \includegraphics[width=0.20\textwidth,clip]{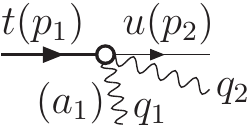} 
\hspace{5mm}
 \includegraphics[width=0.25\textwidth,clip]{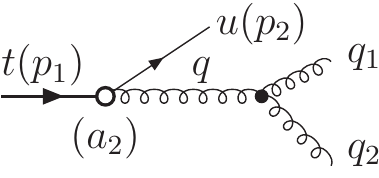} \\[2mm] 
 \includegraphics[width=0.55\textwidth,clip]{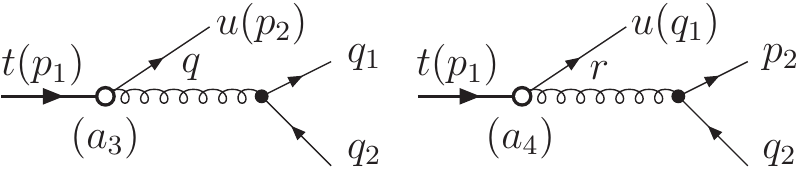} 
\end{center}
  \caption{The diagrams describing the $t$-quark decays.
  Here $ p_1 $ and $ p_2 $ are the momenta of $t$ and $u$-quarks,
respectively, $ q_1 $ and $ q_2 $ are the momenta of the gauge
bosons or $\bar{q} q $ pair.
}
\label{diagr}
\end{figure}

In all calculations we set: $m$ is the mass of the $t$-quark,
the masses of light quarks are assumed to be zero. We use
axial gauge~\cite{itzykson}:
\begin{eqnarray}
  \sum_{pol} V^{\mu} V^{\nu}  = \rho^{\mu \nu}(q) = -g^{\mu \nu}
  + \frac{q^{\mu} n^{\nu} + n^{\mu} q^{\nu}}{(qn)} -
  \frac{n^2 q^{\mu} q^{\nu}}{(qn)^2};  
  \quad \rho^{\mu \nu} n_{\nu} = 0 \label{polar}
 \end{eqnarray}
where $q$ is the gauge boson momentum, 
$n$ is gauge fixing 4-vector. In what follows we take it a sum
of $q_1$ and $q_2$. Then, we get:
\begin{eqnarray*}
   n = q_1 + q_2 &\to&
  \rho^{\mu \nu}(q_1) = \rho^{\mu \nu}(q_2) = \rho^{\mu \nu} =
  -g^{\mu \nu}
  + \frac{q_1^{\mu} q_2^{\nu} + q_2^{\mu} q_1^{\nu}}{(q_1 q_2)} 
  \\
  && \rho^{\mu \nu}\rho^{\alpha \nu} = \rho^{\mu \alpha}; \;\;
  \rho^{\mu \nu} g_{\mu \nu} = 2 \nonumber 
 \end{eqnarray*}
As is well known, the $ 1 \to 3 $ decay is described by two independent
invariants. In this article, the following variables are selected:
\begin{eqnarray}
  &&q^2 = (q_1+q_2)^2; \;\;
 x = \frac{q^2}{m^2}, \;\; y = \frac{(p_2 + q_2)^2}{m^2}
 , \;\; 
  x+y \le 1, \;\;   0 \le \{x,y\} \le 1, \label{variables}  
\end{eqnarray}
Then, we get the following form
for the width for each decay channels:
\begin{eqnarray}
  d\Gamma(t \to u a b) = \frac{m}{96 \pi} |T(t \to u a b)|^2 \, dx \, dy
  \label{dgamma}
  \end{eqnarray}
where $|T|^2 $ is the square of the amplitude (with averaging
on the spin and color of the initial $t$-quark is included in
expression~(\ref{dgamma})), $a$ and $b$ are the corresponding
final states
(photons, gluons, and quarks).\\[2mm]


\noindent {\bf $t\rightarrow u \gamma \gamma$ decay} \\
The amplitude of the decay of the $t$-quark into two photons
is described by a single Feynman diagram ($a_1$ in Fig.~\ref{diagr})
and equals (see Appendix for the vertex $ t \to q \gamma \gamma $):
\begin{eqnarray*} 
  T(t \to u \gamma \gamma)& =& \frac{e_q^2}{\Lambda^3}
  \bar{u}(p_2) (T_1 + T_2 + T_3) u(p_1) \label{dec-gam-gam-1} 
  \end{eqnarray*}
где 
\begin{eqnarray*}
&&  T_1 =  \kappa_1   \;
  \left[(q_1 + q_2)^2 g^{\alpha \beta} - q_2^{\alpha} (q_1 + q_2)^{\beta}
    - (q_1+q_2)^{\alpha} q_1^{\beta}   \right] A_1^{\alpha} A_2^{\beta} 
\\
&& T_2 =  2  \kappa_2  \;
  \left[(q_1 q_2) g^{\alpha \beta} - q_1^{\beta} q_2^{\alpha} 
\right] A_1^{\alpha} A_2^{\beta}
 \\ 
&& 
 T_3 =  2   \,  \kappa_3 \;
 \varepsilon^{\mu \nu \alpha \beta} q_1^{\mu} q_2^{\nu} A_1^{\alpha} A_2^{\beta},
\quad   \kappa_i = \xi^{\gamma}_i + \zeta^{\gamma}_i \gamma^5 
 \end{eqnarray*}
where $q_1$ и $q_2$ are the momenta of the photons.
Using the axial gauge~(\ref{polar}) we get:
\begin{eqnarray*}  
&&q_1^2 = q_2^2 = 0; \quad (q_1 A_1) = (q_1 A_2) = (q_2 A_1) = (q_2 A_2) =0
  \\
&&  T_1 + T_2  = 2(\kappa_1 + \kappa_2) \, B_0; \quad
  T_3 = 2 \kappa_3 \, B_1  \\
 &&   B_0 = (q_1 q_2)( A_1 A_2) = (q^2/2)( A_1 A_2); \quad
  B_1 =  \varepsilon^{\mu \nu \alpha \beta} q_1^{\mu} q_2^{\nu} A_1^{\alpha}
  \\
 && |B_0|^2 = \frac{q^4}{4}   A_1^{\mu} A_2^{\mu} A_1^{\alpha} A_2^{\alpha} =
 \frac{q^4}{4}  \rho^{\mu \alpha}_0 \rho^{\mu \alpha}_0
 = \frac{1}{2} q^4 
 \\
 && |B_1|^2 = \varepsilon^{\mu \nu \alpha \beta} \varepsilon^{\mu' \nu; \alpha' \beta'}
  q_1^{\mu} q_2^{\nu}  q_1^{\mu'} q_2^{\nu'} A_1^{\alpha} A_2^{\beta}
  A_1^{\alpha'} A_2^{\beta'} = 2 (q_1 q_2)^2 = \frac{1}{2} q^4 \\
  &&
  |\bar{u}(p_2) \kappa_i u(p_1)|^2 = 
 \hbox{Tr}
 (\hat{p_1} + m) (\xi^* - \zeta^* \gamma^5) \hat{p_2}
 (\xi + \zeta \gamma^5) =
   2(|\xi|^2 + |\zeta|^2) (m^2 - q^2) 
\end{eqnarray*}
As a result, the amplitude squared equals:
\begin{eqnarray*}
  && |T|^2 = 4 \left(\frac{3}{2}\right)\frac{e_q^4}{\Lambda^6}
  \mathcal{K}_{\gamma \gamma}  
     (m^2 - q^2) q^4;  \\ 
&& \mathcal{K}_{\gamma \gamma} = \left[|\xi^{\gamma}_1 +\xi^{\gamma}_2|^2
  +  |\xi^{\gamma}_3|^2 +  |\zeta^{\gamma}_1 + \zeta^{\gamma}_2|^2 
  +  |\zeta^{\gamma}_3|^2 \right] \nonumber
\end{eqnarray*}
where $3$ is color coefficient, and 2 in the denominator
takes into account the identity of the final photons. 
Below we present the expressions for the $t$-quark decay width:
\renewcommand{\arraystretch}{1.2}
\begin{eqnarray}
  \left.
  \begin{array}{lcl}
  d \Gamma(t\to u \gamma \gamma) / dx \, dy &=&
  e_t^4  \alpha_e^2 \;  \mm 
  \; (m / 16 \pi) 
  \mathcal{K}_{\gamma \gamma} \;   x^2 \, (1  - x)   
 \\
  \;\; \Gamma(t\to u \gamma \gamma) & = &
 e_t^4  \alpha_e^2 \;  \mm 
 \;  ( m / 480 \pi)  \mathcal{K}_{\gamma \gamma}
  \end{array}
  \right\}
  \label{dec-gam-gam-3}
\end{eqnarray}
\renewcommand{\arraystretch}{1.}
where $\alpha_e $ is the fine structure constant and we use the notation 
\vspace{-1mm}
\begin{eqnarray*} 
  \mu^6_{\Lambda} = \left(\frac{m }{\Lambda}\right)^6
  \end{eqnarray*}  \label{mu6}

\noindent
{\bf $t\rightarrow u\, \gamma \, g$ decay} \\
The decay of the $t$-quark into $u$-quark, the photon and the
gluon occurs
due to FCNC interactions induced by a photon and a gluon.
The amplitude is described by a single Feynman diagram
($ a_1 $ in Fig.~\ref{diagr}) and equals:
\begin{eqnarray*} 
  T(t \to q \gamma \, g)& =& \frac{e_q g_s }{\Lambda^3}
  \bar{u}(p_2) (T^{\gamma g}_1 + T^{\gamma g}_2 + T^{\gamma g}_3 ) u(p_1)
  \end{eqnarray*}
where
\begin{eqnarray*}
  T^{\gamma g}_1 &=& \kappa_1  t^a   \;
 \left[((q_1 + q_2)q_1))g^{\alpha \beta} 
  - (q_1+q_2)^{\alpha} q_1^{\beta} \right] A_{1}^{\alpha} B_{2b}^{\beta}  
  \\ 
& +&  \lambda_1  t^a   \;
 \left[((q_1 + q_2)q_2))g^{\alpha \beta} 
  - q_2^{\alpha} (q_1+q_2)^{\beta} \right] A_{1}^{\alpha} B_{2b}^{\beta}  
  \\ 
 T^{\gamma g}_2 &=& (\kappa_2 +  \lambda_2)   
  \left[ (q_1 q_2)g^{\alpha \beta} - q_2^{\alpha} q_1^{\beta} \right]
  A_{1}^{\alpha} B_{2b}^{\beta} \\
 T^{\gamma g}_3 &=&  (\kappa_3 + \lambda_3 + \lambda_4) t^a
 \varepsilon^{\mu \, \nu \alpha \beta} q_1^{\mu} q_2^{\nu}
 A_{1}^{\alpha} B_{2b}^{\beta},
  \quad    \kappa_i = \xi^{\gamma}_i + \zeta^{\gamma}_i \gamma^5, \;\;
   \lambda_i = \xi^{g}_i + \zeta^{g}_i \gamma^5 
\end{eqnarray*}
Expressions
for the decay widths are equal to the corresponding expressions
from~(\ref{dec-gam-gam-3}) with the replacement
\begin{eqnarray}
  \left\{e_t^4 \alpha_e^2  \mathcal{K}_{\gamma \gamma} \right\}_{\gamma \gamma}
  \to \frac{2}{3} \left\{e_t^2 \alpha_e \alpha_s
  \mathcal{K}_{g \gamma}\right\}_{g \, \gamma}
 \label{dec-g-gam-2}
\end{eqnarray}
where $\alpha_s$ is the QCD coupling. Here we have 
\begin{eqnarray*}
  && \mathcal{K}_{g \, \gamma} = |\xi^{\gamma}_1 +\xi^{\gamma}_2 +
    \xi^{g}_1 +\xi^{g}_2|^2
    +|\xi^{\gamma}_3 + \xi^{g}_3 +\xi^{g}_4|^2
  +  |\zeta^{\gamma}_1  +  \zeta^{\gamma}_2 + \zeta^{g}_1 + \zeta^{g}_2|^2 
  +  |\zeta^{\gamma}_3 + \zeta^{g}_3 + \zeta^{g}_4|^2 
    \nonumber 
\end{eqnarray*}

\noindent
{\bf $t\rightarrow u\, g \, g$ decay} \\
The amplitude of the $t$-quark decay into $u$-quark and two gluons
is described by two Feynman diagrams
($a_1$ and $a_2$ in Fig.~\ref{diagr}) and equals:
\begin{eqnarray*} 
  T(t \to u g \, g)& =& \frac{g_s^2 }{\Lambda^3}
  \bar{u}(p_2) (T^{gg}_0 + T^{g  g}_1 + T^{g g}_2 + T^{g g}_3 + T^{g g}_4) u(p_1)
  \end{eqnarray*}
where
\begin{eqnarray*}
 && T^{gg}_0 =  \lambda_1 t^a  \;
 p^{\mu} k^{\nu} (q^{\mu} g^{\nu \alpha} - q^{\nu}g^{\mu \alpha})
\frac{-\rho^{\alpha \alpha'}(q)}{q^2}
 F^{\alpha' \beta \delta}_{abc} B_{b1}^{\beta} B_{c2}^{\delta} 
 \\ 
&& T^{gg}_1 = \lambda_1   t^at^b  \;
 \left[((p_1 q_2) - (p_2 q_1))g^{\alpha \beta} 
  - p_2^{\alpha} p_1^{\beta} + p_1^{\alpha} p_2^{\beta} - q_2^{\alpha} p_1^{\beta} 
 +  p_2^{\alpha} q_1^{\beta} \right] \times  B_{1a}^{\alpha} B_{2b}^{\beta}  
  \\ 
&& + \;\;\; \;\;  \lambda_1  t^bt^a  \;
 \left[((p_1 q_1) - (p_2 q_2))g^{\alpha \beta} 
  + p_2^{\alpha} p_1^{\beta} - p_1^{\alpha} p_2^{\beta} + q_2^{\alpha} p_2^{\beta} 
 -  p_1^{\alpha} q_1^{\beta} \right] \times  B_{1a}^{\alpha} B_{2b}^{\beta}
 \\
&& T^{gg}_2 =  \lambda_2  
 \left(\frac{1}{3} \delta^{ab} + t^k d^{kab} \right)  \;
 \left[(q_1 q_2)g^{\alpha \beta} - q_2^{\alpha} q_1^{\beta} 
   \right] B_{1a}^{\alpha} B_{2b}^{\beta}
 \\
&& T^{gg}_3 =  \lambda_3  \left[
    \left(\frac{1}{3} \delta^{ab} + t^k d^{kab} \right) q_1^{\mu} q_2^{\nu}
 + 2i t^k f^{k ab} k^{\mu} p^{\nu} \right]  \;
 \varepsilon^{\mu \, \nu \alpha \beta}  B_{1a}^{\alpha} B_{2b}^{\beta}
  \\
&& T^{gg}_4 =  \lambda_4  \;
\left(\frac{1}{3} \delta^{ab} + t^k d^{kab} \right) q_1^{\mu} q_2^{\nu}
 \;
 \varepsilon^{\mu \, \nu \alpha \beta}  B_{1a}^{\alpha} B_{2b}^{\beta},
\quad \lambda_i = \xi^{g}_i + \zeta^{g}_i \gamma^5, \;\;
 \end{eqnarray*}
here  $q_1$ и $q_2$ are the photon and gluon momenta.
\begin{eqnarray*}
F^{\alpha' \beta \delta}_{abc} = i g_s f^{abc}
\left[ (-q_1-q)^{\delta} g^{\alpha' \beta}
+  (-q_2 + q_1)^{\alpha'} g^{\delta \beta}
+ (q + q_2)^{\beta} g^{\alpha' \delta} \right];
\quad
 \rho^{\alpha \alpha'}(q) = -g^{\alpha \alpha'} + \frac{q^{\alpha} q^{\alpha'}}{q^2}   
\end{eqnarray*} 
Then we get: 
\begin{eqnarray*}
  && |T(t\to u\, g \, g)|^2 =
  \frac{4g_s^4}{3 \, \Lambda ^6}   
  (m^2 - q^2)
  \left[ 7 \chi_1 q^4 + 3 \chi_2 (m^2 - q^2)^2\right]
  \\
  && \chi_1 = |\xi^{g}_1 - \xi^{g}_2|^2 + |\xi^{g}_3 + \xi^{g}_4|^2
  +  |\zeta^{g}_1 - \zeta^{g}_2|^2  + |\zeta^{g}_3 + \zeta^{g}_4|^2
  \nonumber
  \\
  && \chi_2 = |\xi^{g}_3|^2 + |\zeta^{g}_3|^2 
\end{eqnarray*}
Below we present the expressions for the $t$-quark decay width:
\renewcommand{\arraystretch}{1.2}
\begin{eqnarray}
  \left .
  \begin{array}{lcl}
 d \Gamma(t\to u \, g \, g) / dx \, dy &  = &
  \alpha_s^2 \;  \mm \, (m / 72 \pi) \left[7 \chi_1 x^2(1-x) +
    3 \chi_2 (1-x)^3\right]    
 \\
 \;\;  \Gamma(t\to u \, g  \, g ) & = &
 \alpha_s^2 \;  \mm \; (m / 2160 \pi)
 \left[7 \chi_1  +  18 \chi_2 \right] 
  \end{array}
  \right\} \label{dec-gg-3}
\end{eqnarray}
\renewcommand{\arraystretch}{1.0} \\

\noindent 
{\bf $t\to u \bar{q} q$ and $t \to u \bar{u} u$ decays} \\
The decay of the $t$-quark into a light $u$-quark 
and a quark-antiquark pair
can proceed through two channels
\begin{eqnarray*}
  && t \rightarrow u \,\, \bar{q} \, q, \;\;\; q \ne u; \quad 
 t \rightarrow u \, \bar{u} \, u  
\end{eqnarray*}
The first decay is described by a single diagram $ a_3 $,
and the second decay - by two diagrams $ a_3 $ and $ a_4 $
(see Fig.~\ref{diagr}).
The corresponding amplitudes are:
\begin{eqnarray*}
  && T(t \to u \bar{q} q)_{q\ne u}
  = (g_s^2 / \Lambda^3) \;  W_1;
\;\;  T(t \to u \bar{u} u) 
  = (g_s^2 / \Lambda^3) \; (W_{1} - W_{2}) 
\\ 
&& W_1 = 
  \left\{\bar{u}(p_2) \, \lambda_1 \, t^a  \;  u(p_1)\right\}
      p^{\mu}_1 p^{\nu}_2 (q^{\mu} g^{\nu \alpha} - q^{\nu}g^{\mu \alpha})
  \frac{-\rho^{\alpha' \alpha}(q)}{q^2}
  \left\{\bar{u}(q_1) t^a \gamma^{\alpha'} v(q_2)\right\}
  \\
&& W_2 = \left\{\bar{u}(q_1) \, \lambda_1 \, t^a  \;  u(p_1)\right\}
  p_1^{\mu} q_1^{\nu} (r^{\mu} g^{\nu \alpha} - r^{\nu}g^{\mu \alpha})
    \;\; \frac{-\rho^{\alpha \alpha'}(r)}{r^2}
    \; \left\{\bar{u}(p_2) t^a \gamma^{\alpha'} v(q_2)\right\}  
 \\
 && q = q_1 + q_2; \quad r = p_2 + q_2 
 \\
 &&
 \rho^{\alpha \alpha'}(q) = -g^{\alpha \alpha'} + \frac{q^{\alpha} q^{\alpha'}}{q^2};
  \;\;  \rho^{\alpha \alpha'}(r) =- g^{\alpha \alpha'}
  + \frac{q^{\alpha} r^{\alpha'} + r^{\alpha} q^{\alpha'}  }{(qr) }
 - \frac{q^2 r^{\alpha} r^{\alpha'}}{(qr)^2} 
\end{eqnarray*} 
The expressions for the decay widths of the $t$-quark
are equal to:
\renewcommand{\arraystretch}{1.2}
\begin{eqnarray}
  \left.
  \begin{array}{lcl}
 d \Gamma(t \to  u \, \bar{q} \, q ) / dxdy& =&
  \alpha_s^2 \mm (m / 12 \pi)  \;
  (|\xi^g_1|^2 + |\zeta^g_1|^2) (1-x-y)(1-x)y 
  \\
 \;\;    \Gamma(t \to  u \, \bar{q} \, q ) & = &
  \alpha_s^2 \mm (m / 360 \pi)  \; 
  (|\xi^g_1|^2 + |\zeta^g_1|^2)
  \\ \\
 d \Gamma(t \to  u \bar{u} u) \, dx dy & = & 
  \alpha_s^2 \mm (m / 24 \pi)  \; 
  (|\xi^g_1|^2 + |\zeta^g_1|^2) (1-x-y)(x+y -\frac{7}{3}xy) 
  \\
 \;\;   \Gamma(t \to u \bar{u} u) &=&
  \alpha_s^2 \mm (23 m / 8640 \pi)  \;
  (|\xi^g_1|^2 + |\zeta^g_1|^2)
\end{array}
\right\}
  \label{dec-qq-2}
\end{eqnarray}
\renewcommand{\arraystretch}{1.0}

\vspace{3mm}
\noindent {\bf Estimation of the probability
  the $t$-quark ``two-body'' decays} \\
We note that during hadronization the pair of quarks
(or quark and gluon) with a small invariant mass can
form a hadronic jet ($j$).
In this case, the decays of the $t$-quark in the
experiment can lead to the observed two-body final states:   
\begin{eqnarray*}
  t \to u \gamma \gamma \to j(u\gamma) + \gamma; \;\; 
   u \gamma g \to j(u g) + \gamma;  \;\; 
   u \gamma g \to j(u \gamma) + j(g);  \;\;
   u g g \to j(ug) + j(g), ...
\end{eqnarray*}  
To estimate the probability of these two-body decays 
 we require that the invariant mass of a pair of final 
  particles from the decays~(\ref{channels})
  should be less than 40~GeV
(naturally, the realistic estimates can be obtained after a detailed
modeling of processes):
\begin{eqnarray*}
  m_{min} \le 40 \; \hbox{GeV} \to \delta =
  \left(\frac{m_{min}}{m}\right)^2 \simeq 0.05
  \end{eqnarray*}
We define the probability of the two-body (two-jet) decays as follows:
\begin{eqnarray}
  \beta[t \to jj] = \Gamma(t \to jj)/\Gamma(t \to u ab)
  \label{br1}
\end{eqnarray}
where $a$ and $b$ are two photons, gluons or light quarks.
Then, for decays~(\ref{channels}) we get: 
\begin{eqnarray}
  \left.
  \begin{array}{llll}
     t \to u \gamma \gamma: & 
  \beta[t \to j(u\gamma) + \gamma] & = (5/2) \delta & \simeq 0.13 \\
 t \to u \gamma g: & 
  \beta[t \to j(u\gamma) + j] & = (5/4) \delta +
  (5/2) \delta^3(4-3\delta) & \simeq  0.07 \\
  t \to u g g: & 
  \beta[t \to j + j] & = (5/4) \delta +
  (5/4) (1-(1-\delta)^2) & \simeq  0.3 \\
  t \to u \bar{q} q:  & 
  \beta[t \to j + j] & = 5 \delta(1-\delta) & \simeq  0.24 \\
  t \to u \bar{u} u:  & 
  \beta[t \to j + j] & = (20/23) \delta(6 - 7\delta +
  2\delta^2) & \simeq  0.25
  \end{array} \right\} \label{br2}
  \end{eqnarray}
Thus, it follows from the estimates~(\ref{br2}) that approximately 25\%
the case of decay of $t$-quarks~(\ref{channels}) due to
considered FCNC interaction
can lead to observable two-body final states.

\section {Conclusion}
The contribution of dimension-seven operators to anomalous
FCNC interactions of the
$t$-quarks with a photon and a gluon is considered. 
A phenomenological Lagrangian of such an interaction
and the corresponding Feynman rules are derived. There are evaluated
the expressions for
the widths of the FCNC decays of the $t$-quark to light
$u$ or $c$ quarks and
$\gamma \gamma $, $\gamma \, g$, $\bar{q} q $.
It is shown that in a notable
number of cases such decays of the $t$-quark due to
dimension-seven operators 
can lead to observable two-body (a jet and a photon or two
hadron jets) to final states. \\

In conclusion, the authors are sincerely grateful
V.~Kabachenko, V.~Kachanov, M.~Mangano, P.~Mandrik, A.~Razumov and
R.~Rogalyov for useful discussions.

\section*{Appendix}
\noindent
Here we present the Feynman rules for the anomalous FCNC
interactions of $t$-quarks
due to dimension-seven operators.
All vertices contain the common factor $\Lambda^{- 3} $ and
the notation for anomalous couplings  is used:
\begin{eqnarray*}
  \kappa_i = \xi^{\gamma} + \zeta^{\gamma} \gamma^5;
  \quad
  \lambda_i = \xi^{g} + \zeta^{g} \gamma^5 
\end{eqnarray*}
\noindent
{\bf The interaction with a photon } \\
\begin{tabular}{ll}
  \multirow{3}{*}
{ \includegraphics[width=0.20\textwidth,clip]{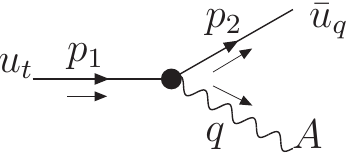} }
 & \\
 & $ 
  e_q  \kappa_{1} \, 
   p_1^{\mu} p_2^{\nu} \left[q^{\mu} g^{\nu \alpha}- q^{\nu} g^{\mu \alpha} \right]
   A^{\alpha} 
   $ \\
   &
   \end{tabular}

\vspace{5mm}

\noindent
\begin{tabular}{ll}
  \multirow{4}{*}
{ \includegraphics[width=0.25\textwidth,clip]{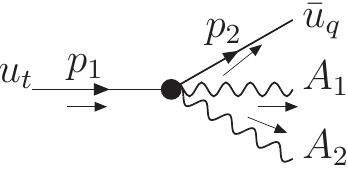}}
&
 $\;\;e_q^2 \,  \kappa_1  
      \;
  \left[(q_1 + q_2)^2 g^{\alpha \beta} - q_2^{\alpha} (q_1 + q_2)^{\beta}
    - (q_1 + q_2)^{\alpha} q_1^{\beta}   \right] A_1^{\alpha} A_2^{\beta} $ \\
  &
 $  2 e_q^2 \,  \kappa_2   \;
  \left[(q_1 q_2) g^{\alpha \beta} - q_1^{\beta} q_2^{\alpha} 
    \right] A_1^{\alpha} A_2^{\beta} $
\\
 &  $ 2  e_q^2 \,  \kappa_3   \;
  \varepsilon^{\mu \nu \alpha \beta} q_1^{\mu} q_2^{\nu} A_1^{\alpha} A_2^{\beta}
  $
\end{tabular}

\vspace{5mm}
\noindent
\begin{tabular}{ll}
  \multirow{4}{*}
{ \includegraphics[width=0.25\textwidth,clip]{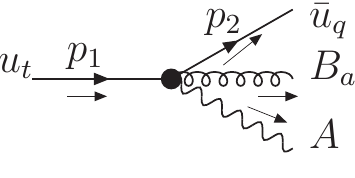} }
&
$e_q g_s \, \kappa_1 t^a   \;
  \left[ \left( (q_1+ q_2)q_1 \right)g^{\alpha \beta}
    - (q_1 + q_2)^{\alpha}q_1^{\beta} \right] 
    A^{\alpha}_1 B_{2\, a}^{\beta} $
\\
& $e_q g_s \,  \kappa_2 t^a  \;
 \, 
  \left[ (q_1q_2)g^{\alpha \beta} - q_1^{\beta} q_2^{\alpha} \right]
  A^{\alpha}_1 B_{2\, a}^{\beta}$
  \\
&
$ e_q g_s \, \kappa_3   t^a \;
 \varepsilon^{\mu \nu \alpha \beta} q_1^{\mu} q_2^{\nu} A^{\alpha}_1 B_{2\, a}^{\beta}
 $
\end{tabular}

\vspace{5mm}
\noindent 
{\bf The interaction with a gluon }

\noindent
\begin{tabular}{ll}
  \multirow{3}{*}
{ \includegraphics[width=0.20\textwidth,clip]{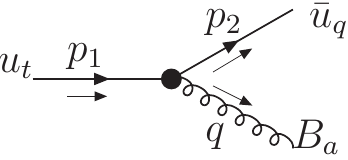} }
 & \\
 & $  g_s \lambda_1 \;
   p_1^{\mu} p_2^{\nu} \left[q^{\mu} g^{\nu \alpha}- q^{\nu} g^{\mu \alpha} \right]
   B_a^{\alpha} 
   $ \\
   &
 \end{tabular}

\vspace{5mm}

\noindent 
\begin{tabular}{ll}
  \multirow{6}{*}
 { \includegraphics[width=0.20\textwidth,clip]{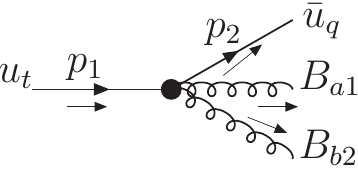} }
           & $g_s^2 W_{gg}^{ab} \; B_{a1}^{\alpha} B_{b2}^{\beta}; \quad
  W_{gg}^{ab} = $\\
 & 
$ \lambda_1  \;  t^a t^a  \;
 \left[((p_1 q_2) - (p_2 q_1))g^{\alpha \beta} 
  - p_2^{\alpha} p_1^{\beta} + p_1^{\alpha} p_2^{\beta} - q_2^{\alpha} p_1^{\beta} 
 +  p_2^{\alpha} q_1^{\beta} \right]  $
 \\
&$  \lambda_1 \; t^b t^a \; 
 \left[((p_1 q_1) - (p_2 q_2))g^{\alpha \beta} 
  + p_2^{\alpha} p_1^{\beta} - p_1^{\alpha} p_2^{\beta} + q_2^{\alpha} p_2^{\beta} 
 -  p_1^{\alpha} q_1^{\beta} \right]$ 
\\
&$  \lambda_2 \; 
 \left( \delta^{ab}/3 + t^k d^{kab} \right)  \;
 \left[(q_1 q_2)g^{\alpha \beta} - q_2^{\alpha} q_1^{\beta} 
 \right] $
 \\
&$ \lambda_3  \;
  \left[
    \left(  \delta^{ab} /3 + t^k d^{kab} \right) q_1^{\mu} q_2^{\nu}
    + 2i t^k f^{k ab} p_2^{\mu} p_1^{\nu} \right]
  \varepsilon^{\mu \, \nu \alpha \beta}  
$
  \\
&$   \lambda_4 \;
  \left(  \delta^{ab} /3 + t^k d^{kab} \right)
  \varepsilon^{\mu \, \nu \alpha \beta} q_1^{\mu} q_2^{\nu}
 $
\end{tabular}

\vspace{5mm}
\noindent
\begin{tabular}{ll}
  \multirow{3}{*}
{ \includegraphics[width=0.25\textwidth,clip]{figs/fig_A_3.pdf} }
&
$e_q g_s  
  \, \lambda_1 \quad \quad \quad t^a  \,
  \left[ ((q_1+q_2)q_2) g^{\alpha \beta}q_2^{\alpha}
    - q_2^{\alpha} (q_1+q_2)^{\beta}  \right]
A_1^{\alpha} B_{2\, a}^{\beta} $
   \\ 
&$  e_q g_s \lambda_2 \quad \quad \quad t^a \,
 \left((q_2 q_1) g^{\alpha \beta} - q_2^{\alpha} q_1^{\beta} \right) 
 A_1^{\alpha} B_{2\, b}^{\beta}$
\\ 
&$ 
e_q g_s 
( \lambda_3 + \lambda_4 )  t^a  \;
 \varepsilon^{\mu \, \nu \alpha \beta} q_1^{\mu} q_2^{\nu} 
A_1^{\alpha} B_{2\, a}^{\beta} $
\end{tabular}

\end{document}